\documentclass[conference,comsoc]{IEEEtran}
\IEEEoverridecommandlockouts
\usepackage{cite}
\usepackage{amsmath,amssymb,amsfonts}
\usepackage{algorithmic}
\usepackage{graphicx}
\usepackage{textcomp}
\usepackage{xcolor}

\newtheorem{theorem}{Theorem}

\begin{document}

\title{Downlink Spectral Efficiency of Massive MIMO with Dual-Polarized Antennas }

\author{\IEEEauthorblockN{\"Ozgecan \"Ozdogan$^*$, Emil Bj\"ornson$^{*,\ddagger}$}
	\IEEEauthorblockA{$^*$Department of Electrical Engineering (ISY), Link\"oping University, Sweden\\
$^\ddagger$Department of Computer Science, KTH Royal Institute of Technology, Sweden\\
		Email: ozgecan.ozdogan@liu.se, emilbjo@kth.se}
		\thanks{This paper was supported by ELLIIT and the Grant 2019-05068 from the Swedish Research Council.}}

\maketitle

\begin{abstract}
This paper considers the downlink of a single-cell massive MIMO (multiple-input multiple-output) system with dual-polarized antennas at both the base station and users. We consider a channel model that takes into account several practical aspects that arise when utilizing dual polarization, such as channel cross-polar discrimination (XPD) and  cross-polar receive and transmit correlations (XPC). We derive the statistical properties of the minimum mean squared error (MMSE) channel estimator for this model. Using these estimates for maximum ratio precoding, a rigorous closed-form downlink spectral efficiency (SE) expression is derived. We compare the SEs achieved in dual-polarized and uni-polarized setups numerically and evaluate the impact of XPD on the downlink SE. 
\end{abstract}


\section{Introduction}

Massive MIMO (multiple-input multiple-output) is the key technology for increasing the spectral efficiency (SE)  in 5G and beyond-5G cellular networks, by virtue of adaptive beamforming and spatial multiplexing \cite{Larsson2014a}. A massive MIMO base station (BS) is equipped with a large number of individually controllable antenna-integrated radios, which can be effectively used to serve tens of user equipements (UEs) simultaneously on the same time-frequency resource. Wireless signals are polarized electromagnetic waves and there exist two orthogonal polarization dimensions.
Practical BSs and UEs typically utilize  dual-polarized antennas (i.e., two co-located antennas that respond to orthogonal polarizations) to squeeze in twice the number of antennas in the same physical enclosure \cite{Asplund20}, as well as capturing signal components from both dimensions.
However, the main theory for massive MIMO has been developed for uni-polarized single-antenna users \cite{massivemimobook}.

The channel modeling for dual-polarized channels is substantially more complicated than for conventional uni-polarized channels.
Several measurements and channel models considering dual-polarized antennas are reported in prior literature. In \cite{Shafi2006a, Calcev2007}, the authors provide geometry-based channel models based on measurement campaigns for dual-polarized small-scale  MIMO systems. In addition,  \cite{coldrey2008modeling, oestges2008dual}  provide analytical channel models based on  extensive surveys of experimental results for single-user dual-polarized MIMO systems. In this paper, we generalize  the widely used model in \cite{coldrey2008modeling} to a massive MIMO setup with multiple UEs. We select this model since it is analytically tractable, yet it includes the effects of channel cross-polar discrimination (XPD),  cross-polar receive and transmit correlations (XPC), and antenna polarization coupling commonly observed in measurements.

The capacity loss due to the polarization mismatch in a single-user dual-polarized MISO system is analyzed in \cite{joung2014capacity}. In \cite{park2015multi}, a scenario with a massive MIMO BS with dual-polarized antennas and multiple users, each equipped with a uni-polarized single antenna, 
is considered. The system operates in frequency division duplex (FDD) mode and the users are grouped based on their spatial correlation matrices. The energy efficiency of a setup similar to \cite{park2015multi} is evaluated in \cite{Yin2020}, while polarization leakage between the antennas is ignored (which makes the channels with different polarizations orthogonal). A distributed FDD massive MIMO system, where each user and  distributed antenna port have a single uni-polarized antenna, is considered in \cite{Park2020}. A multi-user massive MIMO system with non-orthogonal multiple access is considered in \cite{Sena2019}. The users are grouped and it is assumed that the users that are in the same cluster share the same spatial correlation matrix. Some other recent papers related to dual-polarized antennas are \cite{Hemadeh2020,Zhang2020p,Khalilsarai2020}.  In \cite{Hemadeh2020} and \cite{Zhang2020p},  polarization-based modulation schemes are proposed. Furthermore, reconfigurable dual-polarized antennas that are able to change their polarization states are considered in \cite{Hemadeh2020}.  The authors in \cite{Khalilsarai2020} consider the channel correlation matrix estimation problem for dual-polarized antennas.

The canonical form of massive MIMO operates in time-division duplex (TDD) mode and acquires channel state information (CSI) for downlink transmission by using uplink 
 pilot signaling and uplink-downlink channel reciprocity \cite{massivemimobook}. 
 This paper evaluates the performance of a single-cell massive MIMO network with multiple users operating in TDD mode. Different from the majority of previous works, both the BS and UEs are equipped with co-located dual-polarized antennas (as is common in practice). The main contributions are:
\begin{itemize}
	\item[•]  We consider a multi-user massive MIMO scenario with dual-polarized antennas at both the BS and UE sides, and spatial correlation at both sides. To the best of our knowledge, this case has only been studied with equal transmit spatial correlation matrices among the users \cite{Sena2019}.
	
	\item[•]  We derive the minimum mean square error (MMSE) channel estimator and 
	characterize its statistics. Using the estimates for MR
	precoding, we compute a rigorous closed-form downlink SE.
	
	\item[•]  We compare the SEs with dual-polarized and uni-polarized antenna setups numerically, considering both MR  and zero forcing (ZF) precoding.
\end{itemize}

\section{System model with dual-polarized antennas}

We consider a single-cell massive MIMO system with  $\frac{M}{2}$  dual-polarized  antennas at the BS and $K$ UEs, each equipped with a single dual-polarized antenna.  Each dual-polarized antenna is composed of one vertical (V) and one horizontal (H) polarized antennas that are co-located.\footnote{The analysis holds for any set of two orthogonally polarization directions. We just refer to them as V and H polarized for notational convenience.} A V/H polarized antenna emits and receives electromagnetic waves whose electric field oscillates in the V/H plane. Thus, the BS has $M$ antennas in total and each UE has two antennas. Note that an array with a given aperture can accommodate twice as many antennas if dual-polarized antennas are utilized, compared to uni-polarized antennas. 
The system operates in TDD mode and we consider the standard block fading model \cite{massivemimobook}, where the channels are static and frequency-flat within a coherence time-frequency block, and varies independently between blocks. We let $\tau_c$ denote the number of transmission samples per block.

Extending  \cite{coldrey2008modeling} and \cite{joung2014capacity} to  $\frac{M}{2}$ dual-polarized antennas and multiple UEs, the propagation channel of the UE $k$ is 
\begin{align}\label{eq4} \nonumber
	&\mathbf{Z}_k = \begin{bmatrix}
		\mathbf{Z}_{k1} &
		\mathbf{Z}_{k2} &
		\hdots&
		\mathbf{Z}_{k\frac{M}{2}} 
	\end{bmatrix} \in \mathbb{C}^{2 \times M}
	\\ &= \begin{bmatrix}
		z_{kV,1V} & \!\!\!z_{kV,1H} &\!\!\! z_{kV,2V} &\!\!\! z_{kV,2H} & \!\!\hdots\!\! &z_{kV,\frac{M}{2}V} & \!\!\!z_{kV,\frac{M}{2}H}\\
		z_{kH,1V} &\!\!\! z_{kH,1H} &\!\!\! z_{kH,2V} &\!\!\! z_{kH,2H} &\!\! \hdots\!\!& z_{kH,\frac{M}{2}V} & \!\!\!z_{kH,\frac{M}{2}H} 
	\end{bmatrix}  
\end{align}
where $z_{kX,mY}$ is the channel coefficient between the $X$ polarized component of the $k$th UE's dual-polarized antenna   and $Y$ polarized component of the $m$th  dual-polarized BS antenna  with $m \in 1, \dots, \frac{M}{2}$, $k \in 1, \dots, K$ and $X, Y \in \left\lbrace V, H \right\rbrace $. Therefore, each block $\mathbf{Z}_{km}\in \mathbb{C}^{2 \times 2}$ describes the relation from $V$ to $V$, $V$ to $H$, $H$ to $H$ and $H$ to $V$ polarized waves.

In free-space, the cross-polar transmissions (e.g., from a V polarized BS antenna to H polarized UE antenna) is zero under ideal conditions. In a practical scenario, the propagation environment causes cross-polarization scattering that changes the initial polarization state of the electromagnetic waves on the way from the transmitter to the receiver. The channel cross-polarization discrimination (XPD) is the channel's ability to maintain radiated or received polarization purity between H and V polarized signals. 
We assume that it is independent of the BS antenna number $m$ and define it for UE $k$ as
\begin{equation}\label{eq3}
	\mathrm{XPD}_k = \frac{ \mathbb{E}\left\lbrace \left| {z}_{kV,mV}\right|^2\right\rbrace }{\mathbb{E}\left\lbrace \left| {z}_{kH,mV}\right|^2\right\rbrace } = \frac{ \mathbb{E}\left\lbrace \left| {z}_{kH,mH}\right|^2\right\rbrace }{\mathbb{E}\left\lbrace \left| {z}_{kV,mH}\right|^2\right\rbrace }=  \frac{1 - q_k}{q_k}
\end{equation}
with $ 0\leq q_k \leq 1$. By introducing this coefficient, we obtain
\begin{align}
	&\mathbb{E}\left\lbrace \left| {z}_{kV,mV}\right|^2\right\rbrace  =  \mathbb{E}\left\lbrace \left| {z}_{kH,mH}\right|^2\right\rbrace = \beta_k \left( 1 - q_k\right) \label{eq1}\\
	&\mathbb{E}\left\lbrace \left| {z}_{kH,mV}\right|^2\right\rbrace  =  \mathbb{E}\left\lbrace \left| {z}_{kV,mH}\right|^2\right\rbrace = \beta_k  q_k \label{eq2}
\end{align}
where $\beta_k$ is the pathloss parameter of UE $k$. Small values of $q_k$ (i.e., high channel XPD) are typically encountered in line-of-sight-dominated outdoor scenarios whereas low channel XPDs are observed in dense scattering environments. It is important to note  that \eqref{eq3} is  different from the average ratio between the instantaneous values, i.e., 
\begin{equation}
\frac{ \mathbb{E}\left\lbrace \left| {z}_{kV,mV}\right|^2\right\rbrace }{\mathbb{E}\left\lbrace \left| {z}_{kH,mV}\right|^2\right\rbrace } \neq \mathbb{E}\left\lbrace \frac{ \left| {z}_{kV,mV}\right|^2 }{ \left| {z}_{kH,mV}\right|^2} \right\rbrace.
\end{equation}
The instantaneous variation between $\left| {z}_{kV,mV}\right|^2$ and $\left| {z}_{kH,mH}\right|^2$ can be as high as $10$ dB due to the polarization selectivity feature of  scattering environments \cite{Asplund2007, oestges2008dual}. 

The polarization correlation matrices that define the correlation between different polarization states  are given as \cite{coldrey2008modeling}
\begin{equation}
	\mathbf{C}_{BS,k} =\begin{bmatrix}
		1 & t_{p,k}\\
		t^*_{p,k} & 1
	\end{bmatrix} \quad \text{and} \quad \mathbf{C}_{UE,k} =\begin{bmatrix}
		1 & r_{p,k}\\
		r^*_{p,k} & 1
	\end{bmatrix}
\end{equation}
where the cross-polar correlation (XPC) terms $t_{p,k}$ and $r_{p,k}$ at the transmitter and receiver side are 
\begin{align}
	&t_{p,k} = \frac{ \mathbb{E}\left\lbrace {z}_{kV,mV}  {z}^*_{kV,mH}\right\rbrace }{\beta_k\sqrt{q_k\left( 1 - q_k\right)} } = \frac{ \mathbb{E}\left\lbrace {z}_{kH,mV}  {z}^*_{kH,mH}\right\rbrace }{\beta_k\sqrt{q_k\left( 1 - q_k\right)} } ,\\
	&r_{p,k} = \frac{ \mathbb{E}\left\lbrace {z}_{kV,mV}  {z}^*_{kH,mV}\right\rbrace }{\beta_k\sqrt{q_k\left( 1 - q_k\right)} } = \frac{ \mathbb{E}\left\lbrace {z}_{kH,mH}  {z}^*_{kV,mH}\right\rbrace }{\beta_k\sqrt{q_k\left( 1 - q_k\right)} }. 
\end{align}
 Thus, each block in \eqref{eq4} can be written as
\begin{equation}
	\mathbf{Z}_{km}= \mathbf{\Sigma}_k \odot \left( \mathbf{C}^\frac{1}{2}_{UE,k}\mathbf{G}_{km}  \mathbf{C}^\frac{1}{2}_{BS,k}\right) 
\end{equation}
where $\odot$ is the Hadamard (element-wise) product, 
\begin{equation}
\!\!\!\!\!	\mathbf{\Sigma}_k = \begin{bmatrix}
		\sqrt{1 - q_k} & \sqrt{ q_k}  \\
		\sqrt{ q_k} & \sqrt{1 - q_k} 
	\end{bmatrix}, 
	 \quad \mathbf{G}_{km} = \begin{bmatrix}
		g_{kV,mV} & g_{kV,mH} \\
		g_{kH,mV} & g_{kH,mH} 
	\end{bmatrix},
\end{equation}
and $\mathbf{G}_{km} $ has i.i.d. circularly symmetric Gaussian entries with $g_{kX,mY} \sim  \mathcal{N}_\mathbb{C}(0, \beta_k)$ for $X,Y \in \left\lbrace V,H\right\rbrace $. 

If we stack the elements related to UE $k$ for different polarization combinations   as $\mathbf{g}_{k,xy}=[g_{kx,1y}, \dots,g_{kx,\frac{M}{2}y} ] \in \mathbb{C}^{\frac{M}{2} \times 1}$ with $x,y \in \left\lbrace V,H\right\rbrace $, then $\mathbf{g}_{k,xy} \sim \mathcal{N}_\mathbb{C}\left( \mathbf{0}, \mathbf{R}_{\mathrm{BS},k} \right) $ with the spatial correlation matrix $\mathbf{R}_{\mathrm{BS},k} \in \mathbb{C}^{\frac{M}{2} \times \frac{M}{2}}$. 
For example, the vector $\mathbf{g}_{k,VH}$ denotes the relation between (without the XPD coefficients) V polarized component of the UE $k$'s antenna and H polarized component of the BS antennas. Since the V and H polarized antennas are co-located, they see the same scattering environment (in the statistical sense) and therefore all these vectors have the same spatial correlation matrix $\mathbf{R}_{\mathrm{BS},k}$, i.e.,  $\mathbf{g}_{k,VV} \sim \mathcal{N}_\mathbb{C}\left( \mathbf{0}, \mathbf{R}_{\mathrm{BS},k} \right) $,  $\mathbf{g}_{k,VH} \sim \mathcal{N}_\mathbb{C}\left( \mathbf{0}, \mathbf{R}_{\mathrm{BS},k} \right) $,  $\mathbf{g}_{k,HV} \sim \mathcal{N}_\mathbb{C}\left( \mathbf{0}, \mathbf{R}_{\mathrm{BS},k} \right) $, and  $\mathbf{g}_{k,HH} \sim \mathcal{N}_\mathbb{C}\left( \mathbf{0}, \mathbf{R}_{\mathrm{BS},k} \right) $. 
Since the two antennas at the UE side are co-located, the receiver spatial correlation matrix is simply $\mathbf{R}_\mathrm{UE,k}=1, \forall k$. Hence, the propagation channel  of the UE $k$ becomes
\begin{align}\label{eq6}
	\!\!\! \mathbf{Z}_k  &=\! \left( \mathbf{1}_{1 \times \frac{M}{2}} \otimes \mathbf{\Sigma}_k  \right)\! \odot\! \left(  \left( \mathbf{R}_\mathrm{UE,k}  \otimes \mathbf{C}_\mathrm{UE,k}\right)^{1/2} \mathbf{S}_k \!\left( \mathbf{R}_{\mathrm{BS},k}  \otimes \mathbf{C}_\mathrm{BS,k}\right)^{1/2}\! \right) 
\end{align}
where $ \mathbf{S}_k\!\! =\begin{bmatrix}
	s_{kV,1V} &\!\!\!\! s_{kV,1H} &\!\! \!\!\!\!\hdots &\!\!\!\! s_{kV,\frac{M}{2}V} &\!\!\!\! s_{kV,\frac{M}{2}H}\\
	s_{kH,1V} & \!\!\! s_{kH,1H} &\!\! \!\!\!\! \hdots&\!\!\!\! s_{kH,\frac{M}{2}V} &\!\!\!\! s_{kH,\frac{M}{2}H} 
\end{bmatrix} = \begin{bmatrix} \mathbf{s}_{kv} \\ \mathbf{s}_{kh}\end{bmatrix}  
\in \mathbb{C}^{2 \times M}$ has i.i.d. entries with $\mathcal{N}_\mathbb{C}(0, 1 )$-distribution. The operator $\otimes$ denotes the Kronecker product and it arises since the transmitter XPC term $t_{p,k}$ is the same for each antenna element $m$ and the receiver XPC coefficient $r_{p,k}$ is equal for $H$ and $V$ antenna elements of UE $k$.

The channel $\mathbf{H}_k \in \mathbb{C}^{2\times M}$ including the BS and UE antenna polarization matrices that model the antenna hardware effects is
\begin{equation}
	\mathbf{H}_k = \mathbf{F}_\mathrm{UE,k}\mathbf{Z}_k \left( \mathbf{I}_\frac{M}{2} \otimes \mathbf{F}_\mathrm{BS} \right) .
\end{equation}
where $\mathbf{F}_\mathrm{UE,k}$  and $\mathbf{F}_\mathrm{BS}$  are the polarization matrices at the UE $k$ and BS, respectively. If there is no coupling between the two orthogonally polarized ports (V/H),  the polarization matrices are modeled by $\mathbf{F}_\mathrm{X} = [\mathbf{f}_1,  \mathbf{f}_2 ]$ where $\mathbf{f}_1$ and $\mathbf{f}_2$ are unit norm and orthogonal to each other i.e., 	$\mathbf{F}_\mathrm{X} = \begin{bmatrix}
	1& 0\\
	0 &1
\end{bmatrix}$. On the other hand, for imperfect antennas, there is a coupling between the antenna elements and then $\mathbf{f}_1$ and $\mathbf{f}_2$ are not orthogonal anymore. The ratio of the power coupled between the V and H antennas of a dual-polarized antenna is determined by the  antenna's cross-polar isolation (XPI) ability. The XPI is a feature of the antennas solely and independent from the propagation channel. Infinite XPI means no coupling whereas finite values of XPI implies that the antenna hardware is imperfect. For simplicity, we assume that the antenna hardware can perfectly isolate the polarization states.\footnote{For well-designed antennas, the XPI will be negligible compared to the channel XPD.} In addition, various measurements indicate that the transmit and receiver XPCs are close to zero, see \cite[Table 3.1]{clerckx2013mimo}. Therefore, we assume that $t_{p,k}=r_{p,k}= 0$, i.e., $\mathbf{C}_{BS,k} =\mathbf{C}_{UE,k} = \mathbf{I}_2$.  Then, we can write
\begin{align}
	\mathbf{H}_k &= \mathbf{F}_\mathrm{UE,k}\mathbf{Z}_k \left( \mathbf{I}_\frac{M}{2} \otimes \mathbf{F}_\mathrm{BS} \right) \nonumber\\
	&=\left( \mathbf{1}_{1 \times \frac{M}{2}} \otimes \mathbf{\Sigma}_k  \right) \odot \left(   \mathbf{S}_k \left( \mathbf{R}_{\mathrm{BS},k}  \otimes \mathbf{I}_{2}\right)^{1/2} \right) \nonumber \\
	&=  \left[ \left( \mathbf{1}_{1 \times \frac{M}{2}} \otimes  \mathbf{\Sigma}_k \right) \odot \mathbf{S}_k \right] \mathbf{R}^{1/2}_k  \nonumber \\
	&=  \begin{bmatrix}
		\tilde{s}_{kV,1V} & \tilde{s}_{kV,1H} &  \hdots &\tilde{s}_{kV,\frac{M}{2}V} & \tilde{s}_{kV,\frac{M}{2}H}\\
		\tilde{s}_{kH,1V} & \tilde{s}_{kH,1H} &  \hdots& \tilde{s}_{kH,\frac{M}{2}V} & \tilde{s}_{kH,\frac{M}{2}H} 
	\end{bmatrix} \mathbf{R}^{1/2}_k  \nonumber \\
&\overset{\Delta}{=} \tilde{\mathbf{S}}_k\mathbf{R}^{1/2}_k 
\end{align}
where $\mathbf{R}_k = \mathbf{R}_{\mathrm{BS},k}  \otimes \mathbf{I}_2$ (also Hermitian) and $ \tilde{\mathbf{S}}_k = \left( \mathbf{1}_{1 \times \frac{M}{2}} \otimes  \mathbf{\Sigma}_k \right) \odot \mathbf{S}_k = \begin{bmatrix}
	\tilde{\mathbf{s}}^H_{kV} \\ \tilde{\mathbf{s}}^H_{kH}
\end{bmatrix}  $. Notice that 
$\mathbb{E}\left\lbrace |\tilde{s}_{kV,mV}|^2 \right\rbrace = \mathbb{E}\left\lbrace | \tilde{s}_{kH,mH}|^2 \right\rbrace = {1 - q_k}$ and $\mathbb{E}\left\lbrace | \tilde{s}_{kV,mH}|^2 \right\rbrace = \mathbb{E}\left\lbrace | \tilde{s}_{kH,mV}|^2 \right\rbrace = { q_k}$ and they have zero mean. Then, $\mathbf{H}_k = \begin{bmatrix}
	\mathbf{h}_{kV} &  \mathbf{h}_{kH}
\end{bmatrix}^H \in \mathbb{C}^{2 \times M}$ where the row vectors show the channels between the V and H components of UE $k$ and  all BS antennas:
\begin{align}
	&\mathbf{h}_{kV} = \mathbf{R}^{1/2}_{k} \tilde{\mathbf{s}}_{kV} \in \mathbb{C}^{M \times 1},\\
	&\mathbf{h}_{kH} = \mathbf{R}^{1/2}_{k} \tilde{\mathbf{s}}_{kH} \in \mathbb{C}^{M \times 1}.
\end{align}
The covariance matrices of these matrices will be utilized during the channel estimation and are computed as
\begin{align}
	\mathbb{E}\left\lbrace \mathbf{h}_{kV} \mathbf{h}^H_{kV}\right\rbrace 
	&= \mathbf{R}^{1/2}_{k} \begin{bmatrix}
		{1 - q_k}  &0& \dots && 0\\
		0 & q_k\\
		\vdots&&\ddots&& \vdots \\
		&&&{1 - q_k}  & 0\\
		0&\dots&& 0& q_k\\
	\end{bmatrix} \left( \mathbf{R}^{1/2}_{k}\right)^H \nonumber\\
& \overset{\Delta}{=}  \mathbf{R}^{v}_{k},
\end{align}
and \vspace{-0.45cm}
\begin{align}
	\mathbb{E}\left\lbrace \mathbf{h}_{kH} \mathbf{h}^H_{kH}\right\rbrace  
	&= \mathbf{R}^{1/2}_{k}\begin{bmatrix}
		q_k  &0& \dots && 0\\
		0 & 1 -q_k\\
		\vdots&&\ddots&& \vdots \\
		&&& q_k  & 0\\
		0&\dots&& 0& 1 -q_k\\
	\end{bmatrix}  \left( \mathbf{R}^{1/2}_{k}\right)^H \nonumber \\
& \overset{\Delta}{=} \mathbf{R}^{h}_{k}
\end{align}
where $\mathbf{R}^{v}_{k} + \mathbf{R}^{h}_{k} = \mathbf{R}_{k}$. Besides, we calculate
\begin{align}\label{eq5}
\!\!\!\!\!\!	\mathbb{E}\left\lbrace \mathrm{vec}\left(\mathbf{H}^H_k \right)\mathrm{vec}\left(\mathbf{H}^H_k \right)^H\right\rbrace \! = \begin{bmatrix}
		\mathbf{R}^{v}_{k} & \mathbf{0}\\
		\mathbf{0} &  \mathbf{R}^{h}_{k}
	\end{bmatrix} \overset{\Delta}{=} \mathbf{R}_{bk} \! \in \mathbb{C}^{2M \times 2M}
\end{align} 
where $\mathrm{vec}(\cdot)$ denotes vectorization. Notice that \eqref{eq5} implies $	\mathbb{E}\left\lbrace \mathbf{h}_{kH} \mathbf{h}^H_{kV}\right\rbrace = 	\mathbb{E}\left\lbrace \mathbf{h}_{kV} \mathbf{h}^H_{kH}\right\rbrace = \mathbf{0}$
 since the receiver side correlation matrix $\left( \mathbf{R}_\mathrm{UE,k}  \otimes \mathbf{C}_\mathrm{UE,k}\right)^{1/2} = \mathbf{I}_2$ and $\mathbf{C}_\mathrm{BS,k}= \mathbf{I}_2$ for the reason that  V and H polarized waves fade independently through the channel \cite{Asplund2007}.

\section{Channel Estimation}
Each BS requires CSI for uplink receive processing and downlink transmit precoding. Therefore, $\tau_p$ samples are reserved for performing uplink pilot-based channel estimation in each coherence block, giving room for $\tau_p$
mutually orthogonal pilot sequences. Following \cite{Li2016b,Bjornson2010a}, each UE sends its pilot signal $\boldsymbol{\Phi}_k \in \mathbb{C}^{2\times \tau_p}$ to the BS with $\tau_p =2K$ (and $2K \leq \tau_c$).
 The pilot signal is designed as $\boldsymbol{\Phi}_k = \mathbf{L}^{1/2}_k \mathbf{V}^T_k$ where $\mathbf{L}_k = \mathrm{diag}\left( p_{kV}, p_{kH}\right) $ is a pilot allocation matrix with $p_{kV}, p_{kH}$ being the pilot powers  allocated to the V and H polarized antennas, respectively.
 The orthogonal pilot matrix $\mathbf{V}_k \in \mathbb{C}^{2\times \tau_p}$ is designed so that $\mathbf{V}^H_k \mathbf{V}_k = \tau_p\mathbf{I}_{2} $ and $\mathbf{V}^H_k \mathbf{V}_l = \mathbf{0}_2 $ if $l \neq k$.  Also, $\mathrm{tr}\left(\boldsymbol{\Phi}_k \boldsymbol{\Phi}^H_k  \right)/ \tau_p \leq P_k $ where $P_k$ is the total uplink pilot power of UE $k$. Thus, 
\begin{equation}
	\boldsymbol{\Phi}_k\mathbf{V}^*_k =  \begin{bmatrix}
		\sqrt{p_{kV}}\tau_p,  & 0\\
		0 & \sqrt{p_{kH}}\tau_p
	\end{bmatrix} =  \tau_p\mathbf{L}^{1/2}_k,
\end{equation}
\begin{equation}
	\boldsymbol{\Phi}_k\mathbf{V}^*_l  = \mathbf{0}_{2}, \quad l \neq k.
\end{equation}

All UEs transmit their pilot signals simultaneously. 
The received pilot signal  $\mathbf{Y}   \in  \mathbb{C}^{M \times \tau_p}$ at the BS is given by 
\begin{equation}
	\mathbf{Y} = \sum_{l=1}^K \mathbf{H}^H_l \boldsymbol{\Phi}_l +\mathbf{N}
\end{equation}
where $\mathrm{vec}\left(\mathbf{N} \right) \sim \mathcal{N}_\mathbb{C}( \mathbf{0}, \sigma^2 \mathbf{I}_{M\tau_p} ) $ is the receiver noise with variance $\sigma^2$. To estimate the channel of UE $k$, the BS can first process the receive signal by correlating it with the UE's pilot signal.
The processed pilot signal $ \mathbf{Y}^p_k  \in \mathbb{C}^{M \times 2}$ is
\begin{equation}\label{eq_processed}
	\mathbf{Y}^p_k =	\mathbf{Y}\mathbf{V}^*_k = \tau_p\mathbf{H}^H_k  \mathbf{L}^{1/2}_k+ \mathbf{N}\mathbf{V}^*_k .
\end{equation}
Vectorizing  \eqref{eq_processed} gives
\begin{equation}\label{eqrec}
	\mathrm{vec}\left(\mathbf{Y}^p_k \right) =\mathbf{A}  \mathrm{vec}\left( \mathbf{H}^H_k \right) + \mathrm{vec}\left(\mathbf{N}\mathbf{V}^*_k \right) 
\end{equation}
where  $\mathbf{A} = \left( \tau_p\mathbf{L}^{1/2}_k \otimes \mathbf{I}_{M}\right) $ and $\mathrm{vec}\left(\mathbf{N}\mathbf{V}^*_k \right)  \sim \mathcal{N}_\mathbb{C}\left( \mathbf{0}, \sigma^2 \tau_p\mathbf{I}_{2M} \right)$. Besides, the received processed pilot signal can be written as $\mathrm{vec}\left(\mathbf{Y}^p_k \right) \overset{\Delta}{=} \begin{bmatrix}
	\mathbf{y}^p_{kV}\\
	\mathbf{y}^p_{kH}
\end{bmatrix} $ where
\begin{align}
\!\!\!	\mathbb{E}\left\lbrace\mathbf{y}^p_{kV} (\mathbf{y}^p_{kV})^H\right\rbrace  = 
	\tau_p\left(p_{kV}\tau_p\mathbf{R}^{v}_{k} + \sigma^2\mathbf{I}_M \right) \overset{\Delta}{=}\tau_p \left( \mathbf{\Psi}^v_k\right)^{-1},
\end{align}
\begin{align}
\!\!\!\!\!	\mathbb{E}\left\lbrace\mathbf{y}^p_{kH} (\mathbf{y}^p_{kH})^H\right\rbrace  = 
	\tau_p\left(p_{kH}\tau_p\mathbf{R}^{h}_{k} + \sigma^2\mathbf{I}_M \right) \overset{\Delta}{=} \tau_p \left( \mathbf{\Psi}^h_k\right)^{-1} .
\end{align}

Then, based on \eqref{eqrec}, the MMSE estimate of $\mathbf{H}_k $ is \cite{Bjornson2010a}
\begin{align}
	\mathrm{vec}\left( \hat{\mathbf{H}}^H_k \right) &= \mathbf{R}_{bk}\mathbf{A}^H \left(\mathbf{A}\mathbf{R}_{bk} \mathbf{A}^H + \sigma^2 \tau_p\mathbf{I}_{2M} \right)^{-1} \mathrm{vec}\left(\mathbf{Y}^p_k \right)\nonumber \\
	&=\begin{bmatrix}
		\sqrt{p_{kV}}\mathbf{R}^{v}_{k} \mathbf{\Psi}^v_k 	\mathbf{y}^p_{kV}\\
		\sqrt{p_{kH}}\mathbf{R}^{h}_{k}\mathbf{\Psi}^h_k 	\mathbf{y}^p_{kH}
	\end{bmatrix} \overset{\Delta}{=} \begin{bmatrix}
		\hat{\mathbf{h}}_{kV} \\\hat{\mathbf{h}}_{kH}
	\end{bmatrix},
\end{align}
where the MMSE estimates associated with V/H antennas are independent random variables:
\begin{align}
&\hat{\mathbf{h}}_{kV} \sim  \mathcal{N}_\mathbb{C}\left( \mathbf{0}, \mathbf{\Gamma}_k^v\right) \\
&\hat{\mathbf{h}}_{kH} \sim  \mathcal{N}_\mathbb{C}\left( \mathbf{0}, \mathbf{\Gamma}_k^h\right)
\end{align}
with $\mathbf{\Gamma}_k^v =	{p_{kV}}\tau_p \mathbf{R}^{v}_{k} \mathbf{\Psi}^v_k 	\mathbf{R}^{v}_{k}$ and $\mathbf{\Gamma}_k^h =	{p_{kH}}\tau_p \mathbf{R}^{h}_{k} \mathbf{\Psi}^h_k 	\mathbf{R}^{h}_{k}$ where $\mathrm{tr}\left( \mathbf{\Gamma}_k^v\right) = \mathrm{tr}\left( \mathbf{\Gamma}_k^h\right)$ for equal pilot powers ${p_{kV}} = {p_{kH}}$. The error covariance matrix is
\begin{align}
	\mathbf{C}_\mathrm{MMSE,k} &= \mathbf{R}_{bk} -\mathbf{R}_{bk}\mathbf{A}^H \left(\mathbf{A}\mathbf{R}_{bk} \mathbf{A}^H + \sigma^2 \tau_p\mathbf{I}_{2M} \right)^{-1} \mathbf{A} \mathbf{R}_{bk} \nonumber\\
	&=\begin{bmatrix}
		\mathbf{R}^{v}_{k} -\mathbf{\Gamma}_k^v& \mathbf{0}\\
		\mathbf{0}&	\mathbf{R}^{h}_{k}- \mathbf{\Gamma}_k^h
	\end{bmatrix}  \overset{\Delta}{=}\begin{bmatrix}
		\mathbf{C}^{v}_{k} & \mathbf{0}\\
		\mathbf{0} & \mathbf{C}^{h}_{k}
	\end{bmatrix}.
\end{align}
Note that the estimates $\hat{\mathbf{h}}_{kV} $ and $\hat{\mathbf{h}}_{kH}$ are uncorrelated since the channels $\mathbf{h}_{kV}$ and $\mathbf{h}_{kH}$ are uncorrelated.

\section{Downlink Transmission}

During downlink transmission, the BS transmits simultaneously to all UEs using precoding.
The transmitted downlink signal is 
\begin{equation}
	\mathbf{x} = \sum_{l=1}^K \mathbf{W}_l \mathbf{d}_l
\end{equation}
where   $\mathbf{d}_l \in \mathbb{C}^{2 \times 1}$ is the transmit signal satisfying $\mathbb{E}\left\lbrace\mathbf{d}_l \mathbf{d}^H_l \right\rbrace = \mathbf{I}_2 $ and $\mathbf{W}_l \in \mathbb{C}^{M \times 2}$ is the downlink precoding matrix such that $\mathrm{tr}\left( \mathbb{E}\left\lbrace \mathbf{W}^H_l\mathbf{W}_l \right\rbrace\right) \leq P^\mathrm{dl}_l $ where the total downlink transmit power for each UE is  $P^\mathrm{dl}_l = \rho_{lV} + \rho_{lH} $ with transmit powers at V/H antennas are $\rho_{lV}$ and $\rho_{lH}$, respectively.

The received signal at UE $k$ is denoted by $\mathbf{y}_k \in \mathbb{C}^{2 \times 1}$ and computed as \vspace{-0.5cm}
\begin{align}
	\mathbf{y}_k &= \mathbf{H}_k \mathbf{x} + \mathbf{n}_k = \mathbf{H}_k \sum_{l=1}^K \mathbf{W}_l \mathbf{d}_l + \mathbf{n}_k \nonumber \\
	&= \mathbf{H}_k \mathbf{W}_k \mathbf{d}_k +\mathbf{H}_k \sum_{\substack{l=1 \\ l\neq k}}^K \mathbf{W}_l \mathbf{d}_l + \mathbf{n}_k 
\end{align}
where $\mathbf{n}_k \sim \mathcal{N}_\mathbb{C}\left( \mathbf{0}, \sigma^2\mathbf{I}_2\right) $ is the receiver noise.  
A lower bound on the achievable SE \cite{Li2016b}   is 
\begin{align} \label{eq:rate-expression}
	\!\!\! \!\! \! R_k = \frac{\tau_c -\tau_p}{\tau_c}\log_2 \mathrm{det}\left(  \mathbf{I}_2 \!+ \left( \mathbb{E}\left\lbrace  \mathbf{H}_k \mathbf{W}_k  \right\rbrace\right)^H \boldsymbol{\Omega}_k  \mathbb{E}\left\lbrace  \mathbf{H}_k \mathbf{W}_k  \right\rbrace \right) 
\end{align}
where 
\begin{align} \nonumber
	& \boldsymbol{\Omega}_k = \\
	&\left(  \mathbb{E}\left\lbrace  \mathbf{H}_k \sum_{l=1}^K \mathbf{W}_l \mathbf{W}^H_l  \mathbf{H}^H_k \right\rbrace + \sigma^2 \mathbf{I}_2  -\mathbb{E}\left\lbrace  \mathbf{H}_k \mathbf{W}_k  \right\rbrace \left( \mathbb{E}\left\lbrace  \mathbf{H}_k \mathbf{W}_k  \right\rbrace\right)^H  \right)^{-1}.
\end{align}

This expression can be computed numerically for any choice of precoding. In the special case of MR precoding, the expectations can be computed in closed form as described in the following theorem.

\begin{theorem}
	If MR precoding with \begin{align}
		\mathbf{W}_k &= \begin{bmatrix}
			\frac{\hat{\mathbf{h}}_{kV}}{\sqrt{\mathbb{E}\left\lbrace \left\| \hat{\mathbf{h}}_{kV}\right\|^2\right\rbrace}}  & \frac{\hat{\mathbf{h}}_{kH}}{\sqrt{\mathbb{E}\left\lbrace \left\| \hat{\mathbf{h}}_{kH}\right\|^2\right\rbrace}} 
		\end{bmatrix}\begin{bmatrix}
			\sqrt{\rho_{kV} }& 0 \\
			0 &\sqrt{\rho_{kH} }
		\end{bmatrix}\nonumber\\
		&= \begin{bmatrix}
			\frac{\sqrt{\rho_{kV} }\hat{\mathbf{h}}_{kV}}{\sqrt{\mathrm{tr}\left(\mathbf{\Gamma}_k^v \right)    }}  & 	\frac{\sqrt{\rho_{kH} }\hat{\mathbf{h}}_{kH}}{\sqrt{\mathrm{tr}\left(\mathbf{\Gamma}_k^h \right)    }}.
		\end{bmatrix}
	\end{align}
	 is used based on the MMSE estimator, then the achievable SE in \eqref{eq:rate-expression} can be computed in closed form and is given in  \eqref{eqrate} at the top of the next page.
	 \begin{figure*}
	\begin{align}\label{eqrate}
		\!\!\! R_k=\frac{\tau_c -\tau_p}{\tau_c}\log_2\left(1 +\frac{	\rho_{kV} \mathrm{tr}\left(\mathbf{\Gamma}_k^v\right) }{\sum_{l=1}^K \rho_{lV} \frac{\mathrm{tr}\left( \mathbf{\Gamma}_l^v  \mathbf{R}^{v}_{k}\right)}{\mathrm{tr}\left( \mathbf{\Gamma}_l^v \right)} +\rho_{lH}\frac{\mathrm{tr}\left( \mathbf{\Gamma}_l^h  \mathbf{R}^{v}_{k}\right)}{\mathrm{tr}\left( \mathbf{\Gamma}_l^h \right)} + \sigma^2}\right) + \frac{\tau_c -\tau_p}{\tau_c} \log_2\left(1 + \frac{	\rho_{kH} \mathrm{tr}\left(\mathbf{\Gamma}_k^h\right) }{\sum_{l=1}^K \rho_{lH} \frac{\mathrm{tr}\left( \mathbf{\Gamma}_l^h  \mathbf{R}^{h}_{k}\right)}{\mathrm{tr}\left( \mathbf{\Gamma}_l^h \right)} +\rho_{lV}\frac{\mathrm{tr}\left( \mathbf{\Gamma}_l^v  \mathbf{R}^{h}_{k}\right)}{\mathrm{tr}\left( \mathbf{\Gamma}_l^v \right)} + \sigma^2} \right) 
	\end{align}

		\hrulefill
		\vspace{-0.5cm}
	\end{figure*}

\end{theorem}
\begin{IEEEproof}
The proof is omitted due to space limitations, but follows from direct computation of the expectations.
\end{IEEEproof}

\begin{figure*}

\begin{align}\label{eqrateSimplified}
	R_k=\frac{\tau_c -\tau_p}{\tau_c}\log_2\left(1 +\frac{	\frac{M}{2}\rho_{kV} p_{kV} \tau_p  \beta^2_k \left( \sum_{l=1}^K  p_{lV} \tau_p \beta_{l} + \sigma^2 \right)^{-1} }{\sum_{l=1}^K \rho_{lV} \beta_{k} +\rho_{lH}\beta_{k} + \sigma^2}\right) + \frac{\tau_c -\tau_p}{\tau_c} \log_2\left(1 + \frac{	\frac{M}{2}\rho_{kH} p_{kH} \tau_p  \beta^2_k \left( \sum_{l=1}^K  p_{lH} \tau_p \beta_{l} + \sigma^2 \right)^{-1} }{\sum_{l=1}^K \rho_{lV} \beta_{k} +\rho_{lH}\beta_{k} + \sigma^2} \right) 
\end{align}
\hrule
\vspace{-0.5cm}
\end{figure*}

\subsection{Simplification for spatially uncorrelated channels and Infinite XPDs }

The simplified SE expression for $ \mathbf{R}_{\mathrm{BS},k} = \beta_k \mathbf{I}_\frac{M}{2}$ and $q_k = 0$ for all $k$ is given in \eqref{eqrateSimplified}. The first and second logarithms correspond to the data streams associated with V/H polarizations, respectively.  Notice that the signal terms in the numerators depend on the channel estimation quality and there is an array gain of $\frac{M}{2}$ for each data stream, equal to the number of antennas per polarization. The denominators contain the interference terms. We observe that the each data stream receives interference from both polarizations. The interference terms do not scale with the number of antennas and are products of the data transmission powers and channel gains. The general SE expression in \eqref{eqrate} can be also interpreted in a similar way.

\section{Numerical Results}
 
In this section, we evaluate the performance of dual-polarized antennas under different channel conditions. We have a single-cell massive MIMO network with $\frac{M}{2}$ dual-polarized antennas and $K=10$ UEs. Each UE is equipped with a single dual-polarized antenna. The UEs are
independently and uniformly distributed within a square of
size $0.5 \times 0.5$ $\mathrm{km}^2$ at distances larger than 15 m from the BS. The location of each UE is used when computing the large-scale fading and nominal angle between the UE and BSs.

The BS is equipped with a ULA with half-wavelength
antenna spacing.  For the spatial correlation matrices,
we consider $N = 6$ scattering clusters and the covariance
matrix of each cluster is modeled by the (approximate)
Gaussian local scattering model \cite[Sec. 2.6]{massivemimobook} such that
\begin{equation} \label{NLoS}
	\left[ \mathbf{R}_{\mathrm{BS},k}\right] _{s,m} = \frac{\beta_{k}}{N} \sum_{n=1}^{N} e^{\jmath \pi  (s-m)\sin({\varphi}_{k,n})}  e^{-\frac{\sigma^2_\varphi}{2}\left(\pi (s-m)\cos({\varphi}_{k,n})\right)^2 } 
\end{equation}
where $\beta_{k}$ is the large-scale fading coefficient and ${\varphi}_{k,n} \sim \mathcal{U}[{\varphi}_{k}-40^\circ, \ {\varphi}_{k} + 40^\circ]$ is the nominal angle of arrival (AoA) for the $n$ cluster. The multipath components of a cluster have Gaussian distributed AoAs, distributed around the nominal AoA with the angular standard deviation (ASD) $\sigma_\varphi=5^\circ$. The large-scale fading coefficient is modeled (in dB) as 
\begin{equation}\label{betak}
	\beta_{k}= -35.3 -37.6  \log_{10}\left( {d_{k}} \right) + F_{k}
\end{equation}
where $d_k$ is the distance between the BS and UE $k$ in meters, $F_{k} \sim \mathcal{N}(  0, \sigma^2_\mathrm{sf} )$ is the shadow fading with $\sigma_\mathrm{sf} = 7$.  

We consider communication over a $20$ MHz channel and the total receiver noise power is $-94$ dBm. Each coherence block consists of $\tau_c$ = $200$ samples and $\tau_p$ = 20 pilots are allocated for the channel estimation. The pilot powers $p_{kV}= p_{kH} = 100$ mW and the downlink transmit powers $\rho_{kV}= \rho_{kH} = 100$ mW for every UE $k$. 

Fig.~1 compares the sum downlink SEs averaged over different UE locations and shadow fading realizations with dual-polarized and uni-polarized antennas. The XPD value is 7 dB for all UEs. We consider MR and ZF precoding for both antenna setups. For the dual-polarized antennas, the ZF precoding matrix is implemented as  
\begin{align}
	\mathbf{W}^\mathrm{ZF}_k &= \begin{bmatrix}
		\frac{[\mathbf{W}_\mathrm{all}]_{kV}}{ \left\| [\mathbf{W}_\mathrm{all}]_{kV}\right\|}  & \frac{[\mathbf{W}_\mathrm{all}]_{kH}}{ \left\| [\mathbf{W}_\mathrm{all}]_{kH}\right\|}
	\end{bmatrix}\begin{bmatrix}
		\sqrt{\rho_{kV} }& 0 \\
		0 &\sqrt{\rho_{kH} }
	\end{bmatrix}
\end{align}
where $\mathbf{W}_\mathrm{all}= \mathbf{H}^H_\mathrm{all} \left( \mathbf{H}_\mathrm{all}\mathbf{H}^H_\mathrm{all}\right)^{-1} $ and   $\mathbf{H}_\mathrm{all}= [\hat{\mathbf{H}}_1, \dots, \hat{\mathbf{H}}_k, \dots  \hat{\mathbf{H}}_K  ]$. In the uni-polarized antenna setup, we implemented the MR and ZF precoding vectors as
\begin{align}
&	\mathbf{W}^\mathrm{uni}_k = 
	\frac{\sqrt{\rho_\mathrm{uni}} \hat{\mathbf{h}}_k}{\sqrt{\mathbb{E}\left\lbrace \left\| \hat{\mathbf{h}}_{k}\right\|^2\right\rbrace}}, \\
&	\mathbf{W}^\mathrm{uni, ZF}_k = 
		\frac{\sqrt{\rho_\mathrm{uni}}[\mathbf{W}_\mathrm{uni,all}]_{k}}{ \left\| [\mathbf{W}_\mathrm{uni, all}]_{k}\right\|} 
\end{align} 
where $\mathbf{W}_\mathrm{uni, all}= \mathbf{H}^H_\mathrm{uni,all} \left( \mathbf{H}_\mathrm{uni,all}\mathbf{H}^H_\mathrm{uni,all}\right)^{-1} $,   $\mathbf{H}_\mathrm{uni,all}= [\hat{\mathbf{h}}_1, \dots, \hat{\mathbf{h}}_k, \dots  \hat{\mathbf{h}}_K  ]$. The precoding vectors are generated based on the MMSE estimates of channels ${\mathbf{h}}_k \sim \mathcal{N}_\mathbb{C}(  \mathbf{0}, \mathbf{R}_{\mathrm{BS},k}  )$ as  $\hat{\mathbf{h}}_k \sim \mathcal{N}_\mathbb{C}(  \mathbf{0},p_\mathrm{uni} \tau_\mathrm{uni,p}\mathbf{R}_{\mathrm{BS},k}  \mathbf{\Psi}_k\mathbf{R}_{\mathrm{BS},k} )$ with $\mathbf{\Psi}_k = \left( p_\mathrm{uni} \tau_\mathrm{uni,p}\mathbf{R}_{\mathrm{BS},k} + \sigma^2 \mathbf{I}_\frac{M}{2}\right)^{-1}  $, see \cite[Sec. 4]{massivemimobook} for the details. To have a fair comparison, we set $p_\mathrm{uni} = \rho_\mathrm{uni} = 200$ mW and $\tau_\mathrm{uni,p} = 10$, so that the total power is constant and the pilot lengths are minimized in both setups.
The SE expressions from \cite[Sec.~4.3]{massivemimobook} are utilized to calculate the downlink SE with MR precoding for uni-polarized antennas. 

 By utilizing the two polarization dimensions, the dual-polarized systems can ideally double the multiplexing gain, as the signal-to-noise ratio (SNR) goes to infinity. 
In our setup, we observe that the ratios between the average SEs of the dual-polarized and uni-polarized setups are $1.6$ for ZF precoding and $1.7$ MR precoding. Note that the ratio is not equal to 2 because the XPD is finite (meaning that there is a polarization leakage), the SNR is finite, and the prelog factors  $(\tau_c- \tau_p )/\tau_c$ and $(\tau_c- \tau_\mathrm{uni,p} )/\tau_c$ are different since half the numbers of pilots are used to estimate the uni-polarized channels. The fact that the markers overlap with the curves confirms the validity of our analytical results. 

Fig.~2 shows the cumulative distribution function (CDF)
of downlink SE per UE  for the dual-polarized and uni-polarized antenna configurations with ZF and MR precoders for $M =100$.  The randomness is the UE locations and shadow fading realizations. The UEs with good channels benefit from the polarization multiplexing. Fig.~3 shows the average sum downlink SE of dual-polarized antenna setup for different XPD values with MR and ZF precoding matrices. The same XPD values are used across UEs.  We observe that the SEs are higher when the polarization leakage is zero and MR and ZF precoders are affected by the polarization leakage at similar amounts.

\begin{figure}[h]
  \includegraphics[scale=0.49]{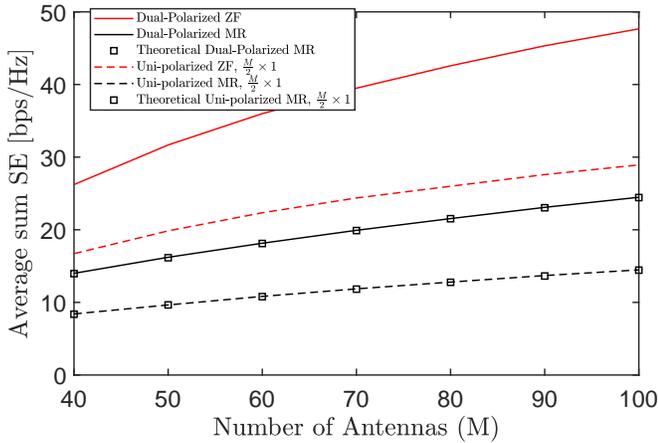} \vspace{-0.3cm}
 \caption{Average downlink sum SE for 10 UEs with different precoders as a function of the number of BS antennas for dual-polarized and uni-polarized setups.}\vspace{-0.4cm}
\end{figure}
	\begin{figure}[h]
	\includegraphics[scale=0.49]{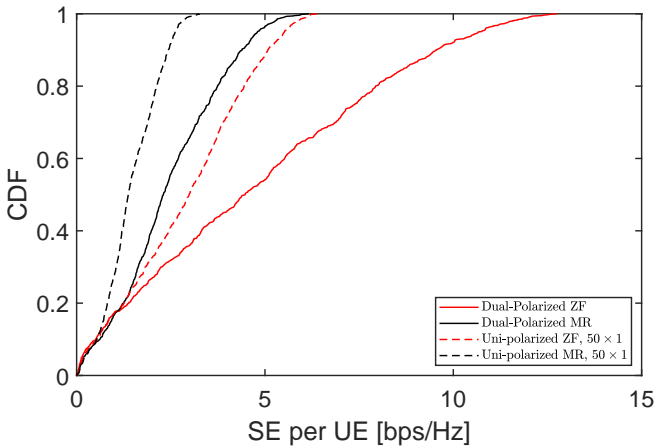} \vspace{-0.3cm}
	\caption{CDF of the downlink SE per UE with $M = 100$ for dual-polarized and uni-polarized setups for different precoders.} \vspace{-0.5cm}
\end{figure}

\begin{figure}[t]
	\includegraphics[scale=0.49]{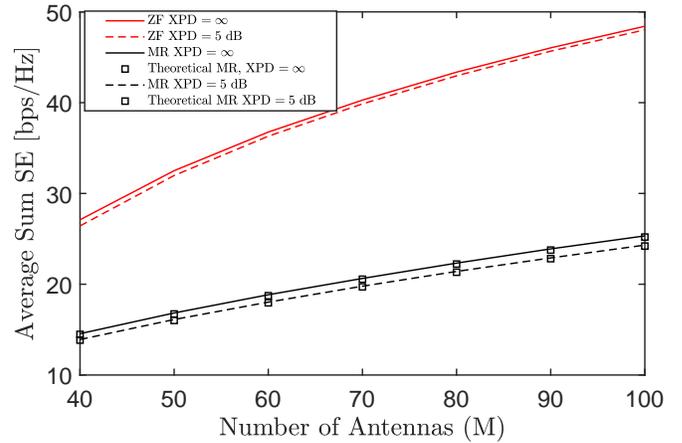} \vspace{-0.3cm}
 \caption{Average downlink sum SE for 10 UEs with different precoders as a function of the number of BS
	antennas for different XPD values.} 
\end{figure}

\section{Conclusions}

This paper studied the downlink of a single-cell massive MIMO system with dual-polarized antennas at both the BS and UEs. We derived closed-form SE expressions for MR precoding based on MMSE estimates. The expression shows how the multiplexing gain can be doubled by utilizing the polarization domain. We observe that dual-polarized arrays have the same physical size and beamforming gain per polarization as a uni-polarized array with half the number of antennas. Hence, the size can be reduced while maintaining or improving the SE.

\bibliographystyle{IEEEtran}
\bibliography{IEEEabrv,refs}

\end{document}